\documentclass[aps,pra,tightenlines,12pt]{revtex4}

\usepackage{graphicx}
\usepackage{amsmath}
\usepackage{amsfonts}
\usepackage{amssymb}
\usepackage{theorem}

\theorembodyfont{\normalfont\sffamily}
\theoremheaderfont{\normalfont\sffamily\bfseries} \theoremstyle{plain}

\begin{document}

\begin{flushright}
{\normalsize\begin{tabular}{r}
UWThPh-2005-31\\
December 2005\\
\\
\end{tabular}}\\
\end{flushright}


\title{Kaonic Qubits}


\author{Reinhold A. Bertlmann and
Beatrix C. Hiesmayr\footnote{Beatrix.Hiesmayr@univie.ac.at}}
\affiliation{Institut f\"ur
Theoretische Physik, University of Vienna, Boltzmanngasse 5, 1090 Vienna, Austria}

\begin{abstract}

\vspace{0.5cm}

Quantum mechanics can also be tested in high energy physics; in particular, the neutral kaon
system is very well suited. We show that these massive particles can be considered as qubits
---\textit{kaonic} qubits--- in the very same way as spin--$\frac{1}{2}$ particles or polarized
photons. But they also have other important properties, namely they are \textit{instable}
particles and they violate the $CP$ symmetry ($C$\dots charge conjugation, $P$\dots parity). We
consider a Bell inequality and, surprisingly, the premises of local realistic theories require
strict $CP$ conservation, in contradiction to experiment. Furthermore we investigate Bohr's
complementary relation in order to describe the physics of the time evolution of kaons. Finally,
we discuss quantum marking and eraser experiments with kaons, which prove in a new way the
very concept of a quantum eraser.
\end{abstract}


\maketitle

\section{Introduction}

For questioning the peculiarities of quantum theory the systems produced in high energy
accelerators are very well suited. The purpose of this Article is to demonstrate various
similarities of such massive systems with the usually considered photonic systems as well as the
important differences and to show new and different tests of the foundations of quantum mechanics.

We are going to focus here on the neutral kaon system (for an overview see
Ref.~\cite{BertlmannSchladming} and references therein). We show that for this system the
superposition principle is realized in the very same way as for photons or spin--$\frac{1}{2}$
particles and that also entangled two--kaon states can be produced. The time evolution ---kaons
oscillate in time and decay--- and the violation of $CP$ symmetry ($C$\dots charge conjugation
transforming a particle into its anti--particle, $P$\dots parity), which means that the world is
not mirrored into an antiworld, are very specific for this system.

Surprisingly, a Bell inequality can be established whose violation is connected to the violation
of $CP$ symmetry. Here two different concepts meet each other, the  concept of local realism and
the concept of symmetry which is fundamental in particle physics.

Furthermore, the nature of kaons offers us to study other puzzling aspects of quantum theory, in
particular, Bohr's complementary principle in a double--slit--like scenario and the quantum
marking and eraser concepts. For the later there exist experimental setups which are not just
analogous to existing quantum eraser experiments but also new ones since the complementary
observables can be measured in two different ways, via ``active'' and ``passive'' measurements.

\section{Kaons as qubits}\label{K-qubits}

What are K-mesons or simply kaons? They are bound states of quarks and anti-quarks ($q\bar q$),
where the quark ($q = u,d,s$) can be the up, down, or strange quark, and can be summarized as
follows:

\vspace{0.1cm}

\begin{center}
\begin{tabular}{|cccc|}
\hline
$\vphantom{\biggr\lbrace}K$-meson & quarks & $S$ & $I_3$ \\
\hline
$K^+$ & $u\bar s$ & $+1$ & $+1/2$ \\
$K^-$ & $\bar us$ & $-1$ & $-1/2 $\\
$K^0$ & $d\bar s$ & $+1$ & $-1/2$ \\
$\bar K^0$ & $\bar ds$ & $-1$ & $+1/2$ \\
$\;K^0$ particle & $\;\bar K^0$ antiparticle & $\;S$ strangeness & $\;I$ isospin \\
\hline
\end{tabular}
\end{center}

\vspace{0.2cm}

Not just for particle physicists the neutral kaon system is unique, these strange mesons are also
fantastic quantum systems, we could even say they are selected by Nature to demonstrate
fundamental quantum principles such as:
\begin{itemize}
    \item [$\bullet$] superposition principle
    \item [$\bullet$] oscillation and decay property
    \item [$\bullet$] quasi-spin property.\\
\end{itemize}

Their mass is about $497$ MeV and they are pseudoscalars $J^P = 0^- \,$. They interact via strong
interactions which are $S$ conserving and weak interactions which are $S$ violating. It is due to
the weak interactions that the kaons oscillate $K^0\,\longleftrightarrow\,\bar K^0\,$.

\subsection*{Quantum states of kaons}\label{K-quantumstates}

Quantum--mechanically we can describe the kaons in the following way. Kaons are characterized by
their \textit{strangeness} quantum number $+1,-1$
\begin{eqnarray}
S|K^0\rangle = + |K^0\rangle \;, \qquad S|\bar K^0\rangle = - |\bar K^0\rangle \;,
\end{eqnarray}
and the combined operation $CP$ gives
\begin{eqnarray}
CP|K^0\rangle = - |\bar K^0\rangle \;, \qquad CP|\bar K^0\rangle = - |K^0\rangle \;.
\end{eqnarray}
It is straightforward to construct the $CP$ eigenstates
\begin{eqnarray}\label{K1K2}
|K_1^0\rangle = \frac{1}{\sqrt{2}}\big\lbrace |K^0\rangle- |\bar K^0\rangle \big\rbrace \;, \qquad
|K_2^0\rangle = \frac{1}{\sqrt{2}}\big\lbrace |K^0\rangle+ |\bar K^0\rangle \big\rbrace\;,
\end{eqnarray}
a quantum number conserved in strong interactions
\begin{eqnarray}
CP|K_1^0\rangle = + |K_1^0\rangle \;, \qquad CP|K_2^0\rangle = - |K_2^0\rangle \;.
\end{eqnarray}

However, due to weak interactions $CP$ symmetry is \textit{violated} and the kaons decay in
physical states, the short-- and long--lived states, $|K_S\rangle , |K_L\rangle$, which differ
slightly in mass, $\Delta m = m_L - m_S = 3.49 \times 10^{-6}$ eV, but immensely in their
lifetimes and decay modes
\begin{eqnarray}\label{kaonSL}
|K_S\rangle = \frac{1}{N}\big\lbrace p |K^0\rangle-q |\bar K^0\rangle \big\rbrace \;, \qquad
|K_L\rangle = \frac{1}{N}\big\lbrace p |K^0\rangle+q |\bar K^0\rangle \big\rbrace \;.
\end{eqnarray}
The weights $p=1+\varepsilon$, $\,q=1-\varepsilon,\,$ with $N^2=|p|^2+|q|^2$ contain the complex
$CP$ \textit{violating parameter} $\varepsilon$ with $\lvert\varepsilon\rvert\approx10^{-3}$.
$CPT$ \textit{invariance} is assumed ($T \dots$ time reversal). The short--lived K--meson decays
dominantly into $K_S\longrightarrow 2 \pi$ with a width or lifetime $\Gamma^{-1}_S\sim\tau_S =
0.89 \times 10^{-10}$ s and the long--lived K--meson decays dominantly into $K_L\longrightarrow 3
\pi$ with $\Gamma^{-1}_L\sim\tau_L = 5.17 \times 10^{-8}$ s. However, due to $CP$ violation we
observe a small amount $K_L\longrightarrow 2 \pi$, which was measured already in 1964. Note that
$CP$ violation means that there is a difference between a world of matter and a world of
antimatter.\\

In this description the superpositions (\ref{K1K2}) and (\ref{kaonSL}) ---or quite generally any
vector in the 2--dimensional complex Hilbert space of kaons--- represent kaonic qubit states in
analogy to the qubit states in quantum information.

\subsection*{Strangeness oscillation}\label{strangenessoscillation}

$K_S, K_L$ are eigenstates of a non--Hermitian ``effective mass'' Hamiltonian
\begin{equation}\label{hamiltonian}
H \, = \, M - \frac{i}{2} \,\Gamma
\end{equation}
satisfying
\begin{equation}
H \,|K_{S,L}\rangle \; = \; \lambda_{S,L} \,|K_{S,L}\rangle \qquad \textrm{with} \qquad
\lambda_{S,L} \, = \, m_{S,L} - \frac{i}{2} \,\Gamma_{S,L} \;.
\end{equation}
Both mesons $K^0$ and $\bar K^0$ have transitions to common states (due to $CP$ violation)
therefore they mix, that means they \textit{oscillate} between $K^0$ and $\bar K^0$ before
decaying. Since the decaying states evolve ---according to the Wigner--Weisskopf approximation---
exponentially in time
\begin{equation}\label{Wigner--Weisskopf}
| K_{S,L} (t)\rangle \; = \; e^{-i \lambda_{S,L} t} | K_{S,L} \rangle \;,
\end{equation}
the subsequent time evolution for $K^0$ and $\bar K^0$ is given by
\begin{eqnarray}\label{K-time-evolution}
| K^0(t) \rangle = g_{+}(t) | K^0 \rangle  + \frac{q}{p} g_{-}(t) | \bar K^0 \rangle \;, \qquad |
\bar K^0(t) \rangle = \frac{p}{q} g_{-}(t) | K^0 \rangle + g_{+}(t) | \bar K^0 \rangle
\end{eqnarray}
with
\begin{equation}\label{g+-}
g_{\pm}(t) \, = \, \frac{1}{2} \left[ \pm e^{-i \lambda_S t} + e^{-i \lambda_L t} \right] \;.
\end{equation}
Supposing that a $K^0$ beam is produced at $t=0$, e.g. by strong interactions, then the
probability for finding a $K^0$ or $\bar K^0$ in the beam is calculated to be
\begin{eqnarray}
\left| \langle K^0 | K^0(t) \rangle \right|^2 &=& \frac{1}{4} \big\lbrace e^{-\Gamma_S t} +
e^{-\Gamma_L t} + 2 \, e^{-\Gamma t}
\cos(\Delta m t)\big\rbrace \;, \nonumber\\
\left| \langle \bar K^0 | K^0(t) \rangle \right|^2 &=& \frac{1}{4} \frac{|q|^2}{|p|^2} \big\lbrace
e^{-\Gamma_S t} + e^{-\Gamma_L t} - 2 \, e^{-\Gamma t} \cos(\Delta m t)\big\rbrace \, ,
\end{eqnarray}
with $\Delta m=m_L-m_S\,$ and $\,\Gamma = \frac{1}{2}(\Gamma_L+\Gamma_S)\,$.

The $K^0$ beam oscillates with frequency $\Delta m / 2\pi$, where $\Delta m \, \tau_S = 0.47$. The
oscillation is clearly visible at times of the order of a few $\tau_S$, before all $K_S$'s have
died out leaving only the $K_L$'s in the beam. So in a beam which contains only $K^0$ mesons at
the beginning $t=0$ there will occur $\bar K^0$ far from the production source through its
presence in the $K_L$ meson.

\subsection*{Quasi--spin of kaons and analogy to photons}\label{quasispin}

In comparison with spin--$\frac{1}{2}$ particles, or with photons having the polarization
directions V (vertical) and H (horizontal), it is very instructive to characterize the kaons by a
\textit{quasi--spin} (for details see Ref.~\cite{BertlmannHiesmayr2001}). We can regard the two
states $| K^0 \rangle$ and $| \bar K^0 \rangle$ as the quasi--spin states up
$\lvert\uparrow\rangle$ and down $\lvert\downarrow\rangle$ and can express the operators acting in
this quasi--spin space by Pauli matrices. So we identify the strangeness operator $S$ with the
Pauli matrix $\sigma_3$, the $CP$ operator with ($-\sigma_1$) and for describing $CP$ violation we
also  need $\sigma_2$. In fact, the Hamiltonian (\ref{hamiltonian}) then has the form
\begin{equation}
H \, = \, a\cdot \mathbf{1} + \vec b \cdot \vec \sigma \;,
\end{equation}
with
\begin{eqnarray}
b_1 = b \cos \alpha, \quad b_2 = b \sin \alpha, \quad b_3 = 0 \;, \nonumber\\
a = \frac{1}{2}(\lambda_L + \lambda_S), \quad b = \frac{1}{2}(\lambda_L - \lambda_S) \;,
\end{eqnarray}
and the angle $\alpha$ is related to the $CP$ violating parameter $\varepsilon$ by
\begin{equation}
e^{i\alpha} \, = \, \frac{1-\varepsilon}{1+\varepsilon} \;.\\
\end{equation}

\vspace{0.2cm}

Summarizing, we have the following kaonic--photonic analogy:

\vspace{0.2cm}

\begin{center}
\begin{tabular}{c c c}
  \textbf{kaon} & \quad\quad \textbf{quasi--spin} & \quad \textbf{photon} \\
  $\lvert K^0\rangle$ & \quad\quad $\lvert\uparrow\rangle_z$ & \quad $\lvert V\rangle$ \\
  $\lvert\bar K^0\rangle$ & \quad\quad $\lvert\downarrow\rangle_z$ & \quad $\lvert H\rangle$ \\
  $\lvert K_1^0\rangle$ & \quad\quad $\lvert\nwarrow\rangle$ & \quad $\lvert -45^0\rangle =
  \frac{1}{\sqrt{2}}(\lvert V\rangle-\lvert H\rangle)$ \\
  $\lvert K_2^0\rangle$ & \quad\quad $\lvert\nearrow\rangle$ & \quad $\lvert +45^0\rangle =
  \frac{1}{\sqrt{2}}(\lvert V\rangle+\lvert H\rangle)$ \\
  $\lvert K_S\rangle$ & \quad\quad $\lvert\rightarrow\rangle_y$ & \quad $\lvert L\rangle =
  \frac{1}{\sqrt{2}}(\lvert V\rangle-i\lvert H\rangle)$ \\
  $\lvert K_L\rangle$ & \quad\quad $\lvert\leftarrow\rangle_y$ & \quad $\lvert R\rangle =
  \frac{1}{\sqrt{2}}(\lvert V\rangle+i\lvert H\rangle)$ \\
\end{tabular}
\end{center}

A good \textit{optical analogy} to the phenomenon of strangeness oscillation can  be achieved by
using the physical effect of birefringence in optical fibers which lead to the rotation of
polarization directions. Thus $H$ (horizontal) polarized light is rotated after some distance into
$V$ (vertical) polarized light, and so on. On the other hand, the decay of kaons can be simulated
by polarization dependent losses in optical fibres, where one state has lower losses than its
orthogonal state~\cite{GisinGo}.
\\
\\
The description of kaons as qubits reveals close analogies to photons but also deep physical
differences. Kaons oscillate, they are massive, they decay and can be characterized by symmetries
like $CP$. Even though some kaon features, like oscillation and decay, can be mimicked by photon
experiments (see Ref.~\cite{GisinGo}), they are inherently different since they are intrinsic
properties of the kaon given by Nature.

\section{Entangled kaons, Bell inequality -- $CP$ violation}\label{BI}

Having discussed kaons as qubit states and their analogy to photons we consider next two qubit
states. A two qubit system of kaons is a general superposition of the 4 states \mbox{$\{| K^0
\rangle \otimes | K^0 \rangle$}, \mbox{$| K^0 \rangle \otimes | \bar K^0 \rangle ,$} $| \bar K^0
\rangle \otimes | K^0 \rangle , | \bar K^0 \rangle \otimes | \bar K^0 \rangle\}\,$.

\subsection*{Entanglement}\label{entanglement}

Interestingly, also for strange mesons entangled states can be obtained, in analogy to the
entangled spin up and down pairs, or H and V polarized photon pairs. Such states are produced by
$e^+ e^-$--colliders through the reaction $e^+ e^- \to \Phi \to K^0 \bar K^0$, in particular at
DA$\Phi$NE in Frascati, or they are produced in $p\bar p$--collisions, like, e.g., at LEAR at
CERN. There, a $K^0 \bar K^0$ pair is created in a $J^{PC}=1^{--}$ quantum state and thus
antisymmetric under $C$ and $P$, and is described at the time $t=0$ by the entangled state
\begin{eqnarray}\label{entangledK0}
| \psi (t=0) \rangle &=&\frac{1}{\sqrt{2}} \left\{ | K^0 \rangle_l \otimes | \bar K^0
\rangle _r - | \bar K^0 \rangle _l \otimes | K^0 \rangle _r \right\}\,,\nonumber\\
&=& \frac{N_{SL}}{\sqrt{2}}\left\{ | K_S \rangle_l \otimes | K_L \rangle _r - | K_L \rangle _l
\otimes | K_S \rangle _r \right\}\,,
\end{eqnarray}
with $N_{SL}=\frac{N^2}{2pq}$, in complete analogy to the entangled photon case
\begin{eqnarray}\label{entangled-photon}
| \psi \rangle &=&\frac{1}{\sqrt{2}} \left\{ | V \rangle_l \otimes | H
\rangle _r - | H \rangle _l \otimes | V \rangle _r \right\}\,,\nonumber\\
&=& \frac{1}{\sqrt{2}}\left\{ | L \rangle_l \otimes | R \rangle _r - | R \rangle _l \otimes | L
\rangle _r \right\}\,.
\end{eqnarray}
The neutral kaons fly apart and are detected on the left ($l$) and right ($r$) hand side of the
source. Of course, during their propagation the $K^0 \bar K^0$ pairs oscillate and the $K_S, K_L$
states decay. This is an important difference to the case of photons which are stable.

Let us measure at time $t_l$ a $K^0$ meson on the left hand side and at time $t_r$ a $K^0$ or a
$\bar K^0$ on the right hand side then we find an EPR--Bell correlation analogously to the
entangled photon case with polarization V--V or V--H. Assuming for simplicity stable kaons
($\Gamma _S = \Gamma _L = 0$) then we get the following result for the quantum probabilities
\begin{eqnarray}
P(K^0,t_l;K^0,t_r) &=& P(\bar K^0,t_l;\bar K^0,t_r) \; = \; \frac{1}{4}
\big\lbrace 1 - \cos(\Delta m(t_l - t_r))\big\rbrace \;,\nonumber\\
P(K^0,t_l;\bar K^0,t_r) &=& P(\bar K^0,t_l;K^0,t_r) \; = \; \frac{1}{4} \big\lbrace 1 +
\cos(\Delta m(t_l - t_r))\big\rbrace \;,
\end{eqnarray}
which is the analogy to the probabilities of finding simultaneously two entangled photons along
two chosen directions $\vec \alpha$ and $\vec \beta$
\begin{eqnarray}
P(\vec \alpha,V ;\vec \beta,V) &=& P(\vec \alpha,H ;\vec \beta,H) \; = \; \frac{1}{4}
\big\lbrace 1 - \cos 2(\alpha - \beta) \big\rbrace \;,\nonumber\\
P(\vec \alpha,V ;\vec \beta,H) &=& P(\vec \alpha,H;\vec \beta,V) \; = \; \frac{1}{4} \big\lbrace 1
+ \cos 2(\alpha - \beta) \big\rbrace \;.
\end{eqnarray}
Thus we observe a \textit{perfect analogy} between times $\Delta m(t_l - t_r)$ and angles
$2(\alpha - \beta)$.\\

Alternatively, we also can fix the time and vary the quasi--spin of the kaon, which corresponds to
a rotation in quasi--spin space analogously to the rotation of polarization of the photon
\begin{eqnarray}
\lvert k\rangle &=& a\lvert K^0\rangle + b\lvert\bar K^0\rangle \quad \longleftrightarrow \quad
\lvert \alpha, \phi; V\rangle = \cos\alpha \lvert V \rangle +
    \sin\alpha \,e^{i \phi}\lvert H \rangle \;.
\end{eqnarray}
\vspace{0.05cm}

Note that the weights $a, b$ are not independent and not all kaonic superpositions are realized in
Nature in contrast to photons.

Depicting the kaonic--photonic analogy we have:

\vspace{0.15cm}

\begin{center}
\small{\hspace{0.2cm}\textbf{kaon propagation} \hspace{1.38cm} \textbf{photon propagation}\\
\setlength{\unitlength}{1cm}

\vspace{0.45cm}

\begin{picture}(10.5,1)(-1.75,-0.6)
    \put(0.48,0){\vector(-1,0){1.5}}
    \put(0.8,0){\vector(1,0){1.5}}
    \put(0.65,0){\circle{0.3}}
    \put(-1.5,0.15){$K^0/K_S$}
    \put(1.7,0.15){$\bar K^0/K_L$}
    \put(3.9,0.15){$V/L$}
    \put(7.1,0.15){$H/R$}
    \put(-0.1,-0.6){\footnotesize{Bell state}}
    \put(-1.2,-0.5){{\footnotesize{left}}}
    \put(1.9,-0.5){{\footnotesize{right}}}
    \put(3.9,-0.5){{\footnotesize{Alice}}}
    \put(7.2,-0.5){{\footnotesize{Bob}}}
    \put(5.68,0){\vector(-1,0){1.5}}
    \put(6.0,0){\vector(1,0){1.5}}
    \put(5.85,0){\circle{0.3}}
    \put(5.4,-0.6){\footnotesize{Bell state}}
\end{picture}

\vspace{0.25cm}

{\footnotesize{\hspace{-3.2cm}$\bullet$ $K^0\bar K^0$ oscillation  \hspace{2.2cm} $\bullet$ stable
\\ \hspace{-7.35cm} $\bullet$ $K_S$, $K_L$ decay
}}}
\end{center}

\vspace{0.3cm}

\subsection*{Bell inequality}\label{Bell-inequality}

Consequently, for establishing a Bell inequality (BI) for kaons we have the option
\begin{enumerate}
\item[$\bullet$] fixing the quasi--spin --- varying time \item[$\bullet$] varying the quasi--spin
--- fixing time.
\end{enumerate}

In this Article we want to concentrate on a BI for certain quasi--spins (the first option we have
studied in detail in Refs.~\cite{BertlmannHiesmayr2001,BBGH}) and show that its violation is
related to a symmetry violation in particle physics. In Ref.~\cite{Nagata,Unnikrishnan} it was
shown that symmetries quite generally may constrain local realistic theories.

For a BI we need $3$ different ``quasi--spins'' -- the ``Bell angles'' -- and we may choose the
$H$, $S$ and $CP$ eigenstates: $|K_S\rangle\,,|\bar K^0\rangle\,$ and $|K_1^0\rangle \,$.

Denoting the probability of measuring the short--lived state $K_S$ on the left hand side and the
anti--kaon $\bar K^0$ on the right hand side, both at the time $t=0$, by $P(K_S,\bar K^0)$, and
analogously the probabilities $P(K_S,K_1^0)$ and $P(K_1^0,\bar K^0)$ we can easily derive under
the usual hypothesis of Bell's locality the following \textit{Wigner--like Bell inequality}
\cite{Uchiyama,BGH-CP}
\begin{equation}\label{UchiyamaBI}
P(K_S,\bar K^0)\; \leq\; P(K_S,K_1^0) + P(K_1^0,\bar K^0) \;.
\end{equation}
BI (\ref{UchiyamaBI}) is rather formal because it involves the unphysical $CP$--even state $|
K^0_1 \rangle$, but -- and this is now important -- it implies an inequality on a
\textit{physical} quantity, the $CP$ violation parameter. Inserting the quantum amplitudes
\begin{equation}
\langle \bar{K}^0\lvert K_S\rangle=-\frac{q}{N}\,,\;\quad \langle \bar{K}^0\lvert K_1^0\rangle = -
\frac{1}{\sqrt{2}}\,,\;\quad \langle K_S\lvert K_1^0\rangle = \frac{1}{\sqrt{2}N}(p^*+q^*)\,,
\end{equation}
and optimizing the inequality we can convert (\ref{UchiyamaBI}) into an inequality for the complex
kaon transition coefficients $p,q$
\begin{eqnarray}\label{inequalpq}
|\,p\,| &\leq& |\,q\,| \;.
\end{eqnarray}
It's amazing, inequality (\ref{inequalpq}) is \textit{experimentally testable}! How does it work?

\subsection*{Semileptonic decays}\label{sect-semileptonicdecays}

Let us consider the semileptonic decays of the kaons. The strange quark $s$ decays weakly as
constituent of $\bar K^0\,$:

\vspace{0.1cm}

\begin{center}
\includegraphics[height=1cm]{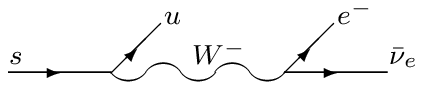}
\end{center}

\vspace{0.3cm}

Due to their quark content the kaon $K^0(\bar s d)$ and the anti--kaon $\bar K^0(s \bar d)$ have
the following definite decays:
\begin{eqnarray}\label{semileptonic-decays}
K^0(d\bar{s}) \;&\longrightarrow&\; \pi^-(d\bar{u})\;\; l^+\;\nu_l \qquad
\textrm{where} \qquad \bar{s} \;\longrightarrow\; \bar{u}\;\; l^+\;\nu_l \nonumber \\
\bar{K}^0(\bar{d}s) \;&\longrightarrow&\; \pi^+(\bar{d}u)\;\; l^-\;\bar{\nu}_l \qquad
\textrm{where} \qquad  s \;\longrightarrow\; u\;\; l^-\;\bar{\nu}_l \;,
\end{eqnarray}
with $l$ either muon or electron, $l=\mu, e\,$. When studying the leptonic charge asymmetry
\begin{eqnarray}\label{asymlept}
\delta &=& \frac{\Gamma(K_L\rightarrow \pi^- l^+ \nu_l) - \Gamma(K_L\rightarrow \pi^+ l^- \bar
\nu_l)}{\Gamma(K_L\rightarrow \pi^- l^+ \nu_l) + \Gamma(K_L\rightarrow \pi^+ l^- \bar \nu_l)} \;,
\end{eqnarray}
we notice that $l^+$ and $l^-$ tag $K^0$ and $\bar K^0$, respectively, in the $K_L$ state, and the
leptonic asymmetry (\ref{asymlept}) is expressed by the probabilities $|p|^2$ and $|q|^2$ of
finding a $K^0$ and a $\bar K^0$, respectively, in the $K_L$ state
\begin{eqnarray}
\delta &=& \frac{|p|^2-|q|^2}{|p|^2+|q|^2} \;.
\end{eqnarray}

Returning to inequality (\ref{inequalpq}) we find consequently the bound
\begin{eqnarray}\label{inequaldelta}
\delta &\leq& 0
\end{eqnarray}
for the leptonic charge asymmetry which measures $CP$ violation.

Experimentally, however, the asymmetry is nonvanishing \cite{ParticleData}
\begin{equation}\label{deltaexp}
\delta = (3.27 \pm 0.12)\cdot 10^{-3} \;.
\end{equation}
What we find is that bound (\ref{inequaldelta}), dictated by BI (\ref{UchiyamaBI}), is in
contradiction to the experimental value (\ref{deltaexp}) which is definitely positive.

On the other hand, we can replace $\bar K^0$ by $K^0$ in the BI (\ref{UchiyamaBI}) and obtain the
reversed inequality $\delta \geq 0$ so that respecting all possible BI's leads to strict equality
$\delta = 0$, $CP$ conservation, in contradiction to experiment.

Thus the experimental fact of $CP$ violation rules out a local realistic theory for the
description of a $K^0 \bar K^0$ system!

\section{Kaons as double slits}\label{doubleslit}

The famous statement ``\textit{the double slit contains the only mystery}'' of Richard Feynman is
well known, his statement about kaons is not less to the point ``\textit{If there is any place
where we have a chance to test the main principles of quantum mechanics in the purest way ---does
the superposition of amplitudes work or doesn't it?--- this is it.}'' \cite{Feynman}. In this
section we argue that single neutral kaons can be considered as double slits as well.

Bohr's complementarity principle or the closely related concept of duality in interferometric or
double slit like devices are at the heart of quantum mechanics. The qualitative well-known
statement that ``\textit{the observation of an interference pattern and the acquisition of
which--way information are mutually exclusive}'' has only recently been rephrased to a
quantitative statement \cite{GreenbergerYasin,Englert}:
\begin{eqnarray}\label{comp}
{\cal P}^2(y)+{\cal V}_0^2(y)\leq 1\;,
\end{eqnarray}
where the equality is valid for pure quantum states and the inequality for mixed ones. ${\cal
V}_0(y)$ is the fringe visibility which quantifies the sharpness or contrast of the interference
pattern (``the wave--like property'') and can depend on an external parameter $y$, whereas ${\cal
P}(y)$ denotes the path predictability, i.e., the \textit{a priori} knowledge one can have on the
path taken by the interfering system (the ``particle--like'' property). It is defined by
\begin{equation}
{\cal P}(y)\;=\;|p_I(y)-p_{II}(y)|\;,
\end{equation}
where $p_I(y)$ and $p_{II}(y)$ are the probabilities for taking each path ($p_I(y)+p_{II}(y)=1)$.
It is often too idealized to assume that the predictability and visibility are independent on an
external parameter. For example, consider a usual double slit experiment, then the intensity is
generally given by
\begin{equation}
I(y)\;=\; F(y)\;\big(1+{\cal V}_0(y) \cos(\phi(y)\big)\;,
\end{equation}
where $F(y)$ is specific for each setup and $\phi(y)$ is the phase--difference between the two
paths. The variable $y$ characterizes in this case the detector position, thus visibility and
predictability are $y$--dependent.

In Ref.~\cite{SBGH3} the authors investigated physical situations for which the expressions of
${\cal V}_0(y), {\cal P}(y)$ and $\phi(y)$ can be calculated analytically. This included
interference patterns of various types of double slit experiments ($y$ is linked to position), but
also oscillations due to particle mixing ($y$ is linked to time), e.g. by the kaon system, and
also Mott scattering experiments of identical particles or nuclei ($y$ is linked to a scattering
angle). All these two--state systems belonging to distinct fields of physics can then be treated
via the generalized complementarity relation in a unified way. Even for specific thermodynamical
systems Bohr's complementarity can manifest itself, see Ref.~\cite{HV}. Here we investigate the
neutral kaon case.

The time evolution of an initial $K^0$ state is given by Eq.(\ref{K-time-evolution}) (in the
following $CP$ violation effects can  safely be neglected)
\begin{eqnarray}
|K^0 (t)\rangle &=& \frac{1}{\sqrt{2}} e^{-i m_L t-\frac{\Gamma_L}{2} t}\left\lbrace e^{i \Delta m
t+\frac{\Delta\Gamma}{2} t}\,  |K_S\rangle +  |K_L\rangle\right\rbrace\;,
\end{eqnarray}
where we denoted $\Delta\Gamma=\Gamma_L-\Gamma_S<0\,$. We are only interested in kaons which
survive up to a certain time $t$, thus we consider the following normalized state
\begin{eqnarray}\label{K-evolution-slit}
\label{singlekaon} |K^0 (t)\rangle &\cong& \frac{1}{\sqrt{2\cosh(\frac{\Delta\Gamma}{2} t)}}\;
e^{-\frac{\Delta\Gamma}{4} t}\left\lbrace e^{i \Delta m t+\frac{\Delta\Gamma}{2} t}\,|K_S\rangle +
|K_L\rangle\right\rbrace\;.
\end{eqnarray}
State (\ref{K-evolution-slit}) can be interpreted as follows. The two mass eigenstates
$|K_S\rangle, |K_L\rangle$ represent the two slits. At time $t=0$ both slits have the same width,
as time evolves one slit decreases as compared to the other, however, in addition the whole double
slit shrinks due to the decay property of the kaons. This analogy gets more obvious if we consider
for an initial $K^0$ the probabilities for finding after a certain time $t$ a $K^0$ or a $\bar
K^0$ state, i.e. the strangeness oscillation
\begin{eqnarray}
P(K^0, t)&=& \left| \langle K^0 | K^0(t) \rangle \right|^2  = \frac{1}{2} \biggl\lbrace
1+\frac{\cos(\Delta m t)}{\cosh(\frac{\Delta\Gamma}{2} t)}\biggr\rbrace\nonumber\\
P(\bar K^0, t)&=& \left| \langle \bar K^0 | K^0(t) \rangle \right|^2 = \frac{1}{2} \biggl\lbrace
1-\frac{\cos(\Delta m t)}{\cosh(\frac{\Delta\Gamma}{2} t)}\biggr\rbrace\;.
\end{eqnarray}
We observe that the oscillating phase is given by $\phi(t)=\Delta m \,t$ and the time dependent
visibility by
\begin{eqnarray}\label{visibility}
{\cal V}_0(t)=\frac{1}{\cosh(\frac{\Delta\Gamma}{2} t)}\;,
\end{eqnarray}
which is maximal at $t=0$. The ``which width'' information corresponding to the path
predictability ${\cal P}(t)$ can be directly calculated from Eq.(\ref{singlekaon})
\begin{eqnarray}\label{predictability}
{\cal P}(t)=\left| P(K_S, t)-P(K_L, t)\right|=\left|\;\frac{e^{\frac{\Delta \Gamma}{2} t} \;-\;
e^{-\frac{\Delta \Gamma}{2}t}}{2\cosh(\frac{\Delta\Gamma}{2} t)}\;
\right|=\left|\;\tanh\big(\frac{\Delta\Gamma}{2} t\big)\right|\;.
\end{eqnarray}
The larger the time $t$ is, the more probable is the propagation of the $K_L$
component, because the $K_S$ component has died out, the predictability
converges to its upper bound $1$.

Both expressions for predictability (\ref{predictability}) and visibility (\ref{visibility})
satisfy the complementary relation (\ref{comp}) for all times $t$
\begin{eqnarray}\label{kaon-complement-relation}
{\cal P}^2(t)+{\cal V}_0^2(t) \;=\; \tanh^2\big(\frac{\Delta\Gamma}{2}t\big) +
\frac{1}{\cosh^2(\frac{\Delta\Gamma}{2}t)} \:=\; 1\;.
\end{eqnarray}
For time $t=0$ there is full interference, the visibility is ${\cal V}_0(t=0)=1$, and we have no
information about the lifetimes or widths, ${\cal P}(t=0)=0$. This corresponds to the usual double
slit scenario. For large times, i.e. $t\gg 1/\Gamma_S$, the kaon is most probable in a long lived
state $K_L$ and no interference is observed, we have information on the ``which width''. For times
between the two extremes we obtain partially information on ``which width'' and on the
interference contrast due to the natural instability of the kaons. However, the full information
on the system is contained in Eq.(\ref{comp}) and is for pure systems always maximal.

The complementarity principle was phrased by Niels Bohr in an attempt to express the most
fundamental difference between classical and quantum physics. According to this principle, and in
sharp contrast to classical physics, in quantum physics we cannot capture all aspects of reality
simultaneously, the information content is always limited. Neutral kaons encapsulate indeed this
peculiar feature in the very same way as a particle travelling through a double slit. But kaons
are double slits provided by Nature for free!

\section{Kaonic quantum eraser}\label{quantumeraser}

Two hundred years ago Thomas Young taught us that photons interfere. Nowadays also experiments
with very massive particles, like the fullerenes, have impressively demonstrated that fundamental
feature of quantum mechanics \cite{Arndt}. It seems that there is no physical reason why not even
heavier particles should interfere except for technical ones\footnote{For example, a ``red Ferrari
racing through a double slit'' (as demonstrated by Markus Arndt at the Symposium ``Bose--Einstein
condensation and quantum information'' at the Erwin Schr\" odinger Institute, Vienna, December
2005)}. In the previous section we have shown that the knowledge on the path through the double
slit is the reason why interference is lost. The gedanken experiment of Scully and Dr\"uhl in 1982
\cite{scully82} surprized the physics community, if the knowledge on the path of the particle is
erased, interference is brought back again.

Since that work many different types of quantum erasures have been analyzed and experiments were
performed with atom interferometers \cite{Duerr} and entangled photons
\cite{Herzog,Kim,Tsegaye,Walborn,Trifonov,KimKim} where the quantum erasure in the so-called
``delayed choice'' mode captures best the essence and the most subtle aspects of the eraser
phenomenon. In this case the meter, the quantum system which carries the mark on the path taken,
is a system spatially separated from the interfering system which is generally called the object
system. The decision \textit{to erase or not} the mark of the meter system ---and therefore
\textit{to observe or not} interference--- can be taken long after the measurement on the object
system has been completed. This was nicely phrased by Aharonov and Zubaiy in their review article
\cite{AharonovZubaiy} as ``erasing the past and impacting the future''.

Here we want to present four different types of quantum erasure concepts for neutral kaons,
Refs.~\cite{SBGH1,SBGH6}. Two of them are analogous to performed erasure experiments with
entangled photons, e.g. Refs.~\cite{Herzog,Kim}. In the first experiment the erasure operation was
carried out ``actively'', i.e., by exerting the free will of the experimenter, whereas in the
latter experiment the erasure operation was carried out ``partially actively'', i.e., the mark of
the meter system was erased or not by a well known probabilistic law, e.g., by a beam splitter.
However, different to photons the kaons can be measured by an \textit{active} or a
\textit{passive} procedure. This offers new quantum erasure possibilities and proves the very
concept of a quantum eraser, namely sorting events.

For neutral kaons there exist two physical alternative bases. The first basis is the strangeness
eigenstate basis $\{| K^0\rangle, |\bar K^0 \rangle\}$, it can be measured by inserting along the
kaon trajectory a piece of ordinary matter. Due to strangeness conservation of the strong
interactions the incoming state is projected either onto $K^0$ by $K^0 p\rightarrow K^+ n$ or onto
$\bar K^0$ by $\bar K^0 p\rightarrow \Lambda \pi^+$, $\bar K^0 n\rightarrow \Lambda \pi^0$ or
$\bar K^0 n\rightarrow K^- p$. Here nucleonic matter plays the same role as a two channel analyzer
for polarized photon beams.

Alternatively, the strangeness content of neutral kaons can be determined by observing their
semileptonic decay modes, Eq.(\ref{semileptonic-decays}).

Obviously, the experimenter has no control of the kaon decay, neither of the mode nor of the time.
The experimenter can only sort at the end of the day all observed events in proper decay modes and
time intervals. We call this procedure opposite to the \textit{active} measurement procedure
described above a \textit{passive} measurement procedure of strangeness.

The second basis $\{K_S,K_L\}$ consists of the short-- and long--lived states having well defined
masses $m_{S(L)}$ and decay widths $\Gamma_{(S)L}$. We have seen that it is the appropriate basis
to discuss the kaon propagation in free space, because these states preserve their own identity in
time, Eq.(\ref{Wigner--Weisskopf}). Due to the huge difference in the decay widths the $K_S$'s
decay much faster than the $K_L$'s. Thus in order to observe if a propagating kaon is a $K_S$ or
$K_L$ at an instant time $t$, one has to detect at which time it subsequently decays. Kaons which
are observed to decay before $\simeq t + 4.8\, \tau_S$ have to be identified as $K_S$'s, while
those surviving after this time are assumed to be $K_L$'s. Misidentifications reduce only to a few
parts in $10^{-3}$, see also Refs.~\cite{SBGH1,SBGH6}. Note that the experimenter doesn't care
about the specific decay mode, he records only a decay event at a certain time. We call this
procedure an \textit{active} measurement of lifetime.

Since the neutral kaon system violates the $CP$ symmetry (recall Section \ref{K-quantumstates})
the mass eigenstates are not strictly orthogonal, $\langle K_S|K_L\rangle\neq 0$. However,
neglecting $CP$ violation ---remember it is of the order of $10^{-3}$--- the $K_S$'s are
identified by a $2\pi$ final state and $K_L$'s by a $3\pi$ final state. We call this procedure a
\textit{passive} measurement of lifetime, since the kaon decay times and decay channels used in
the measurement are entirely determined by the quantum nature of kaons and cannot be in any way
influenced by the experimenter.

\subsection*{(a) Active eraser with \textit{active} measurements}

\begin{figure}
\includegraphics[width=250pt,
keepaspectratio=true]{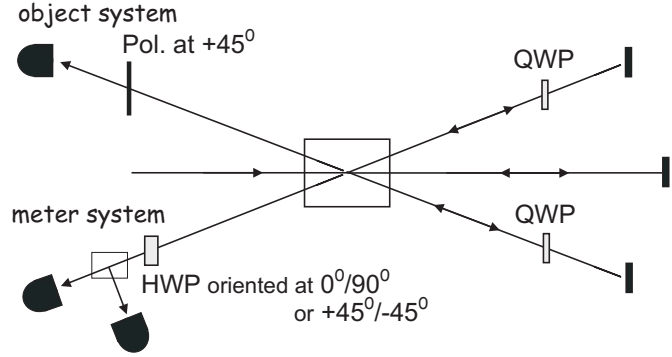} \caption{Here the setup for an active eraser is sketched. A bump
beam transverses twice a e.g. type II crystal. The pairs produced in the first passage through the
crystal cross two times a quarter--wave plate (QWP) which transforms an original horizontal
polarized photon into a vertical one and vice versa. The pairs produced in the second passage
through the crystal is directly directed to the measurement devices. The signal (object) photon is
always measured after crossing a polarization analyzer aligned at $+45^\circ$. The idler (meter)
photon crosses a half--wave plate (HWP) oriented at $0^\circ,90^\circ$ (first setup) or
$\pm45^\circ$ (second setup) and is then analyzed by a polarization beam splitter. In the first
setup
---meter photon is measured in the $H/V$ basis--- one has full \textit{which way} information,
namely if the pair was produced at the first or second passage. In the second
setup ---meter photon is measured in the $+45^\circ/-45^\circ$ basis--- the
information on the first or second passage is erased, one observes fringes or
antifringes.}\label{photonActive}
\end{figure}

Let us first discuss the photon analogy, e.g., the two experimental setups in Ref.~\cite{Herzog}.
In the first setup two interfering two--photon amplitudes are prepared by forcing a pump beam to
cross twice the same nonlinear crystal. Idler and signal photons from the first down conversion
are marked by rotating their polarization by $90^\circ$ and then superposed to the idler (i) and
signal (s) photons emerging from the second passage of the beam through the crystal. If type--II
spontaneous parametric down conversion were used, we had the state \footnote{The authors of
Ref.~\cite{Herzog} used type--I crystals in their experiment.}
\begin{eqnarray}\label{photonentangled}
|\psi\rangle&=& \frac{1}{\sqrt{2}}\biggl\lbrace \underbrace{|V\rangle_i
|H\rangle_s}_{\rm{second\;passage}} - \,e^{i \Delta\phi}\, \underbrace{|H\rangle_i
|V\rangle_s}_{\rm{first\;passage}}\biggr\rbrace\;,
\end{eqnarray}
where the relative phase $\Delta \phi$ is under control by the experimenter (the symbol $\otimes$
for the tensor product of the states is dropped from now on). The signal photon, the object
system, is always measured after crossing a polarization analyzer aligned at $+45^0$, see
Fig.~\ref{photonActive}. Due to entanglement, the vertical or horizontal idler polarization
supplies full \textit{which way} information for the signal (object) system, i.e., whether it was
produced at the first or second passage. No interference can be observed in the signal--idler
joint detections. To erase this information, the idler photon has to be detected in the
$+45^\circ/-45^\circ$ basis.

In case of entangled kaons the state is described by Eq.(\ref{entangledK0}). The analogy with
Eq.(\ref{photonentangled}) is quite obvious, however, kaons evolve in time, such that the state
depends on the two time measurements on the left hand side, $t_l$, and on the right hand side,
$t_r$, or more precise on $\Delta t=t_l-t_r\,$, when normalized\footnote{Thanks to this
normalization, we work with bipartite two--level quantum systems like polarization entangled
photons or entangled spin--$1/2$ particles. For an accurate description of the time evolution of
kaons and its implementation consult Ref.~\cite{BertlmannHiesmayr2001}.} to surviving kaon pairs
\begin{eqnarray}
\label{timeentangled} |\phi(\Delta t)\rangle &=& \frac{1}{\sqrt {1+e^{\Delta\Gamma \Delta
t}}}\biggl\lbrace |K_L\rangle_l|K_S\rangle_r - e^{i \Delta m \Delta t} e^{{1 \over 2} \Delta
\Gamma \Delta t}|K_S\rangle_l|K_L\rangle_r\biggr\rbrace\nonumber\\
&=& \frac{1}{2 \sqrt {1+e^{\Delta\Gamma \Delta t}}} \left\{\big(1-e^{i \Delta m \Delta t} e^{{1
\over 2} \Delta \Gamma \Delta t}\big)
\lbrace|K^0\rangle_l|K^0\rangle_r-|\bar K^0\rangle_l|\bar K^0\rangle_r\rbrace \right. \nonumber\\
&&\hphantom{\frac{1}{2 \sqrt {1+e^{\Delta\Gamma \Delta t}}}} \left. + \big(1+e^{i \Delta m \Delta
t} e^{{1 \over 2} \Delta \Gamma \Delta t}\big) \lbrace |K^0\rangle_l|\bar K^0\rangle_r-|\bar
K^0\rangle_l|K^0\rangle_r\rbrace \right\}\;.
\end{eqnarray}
We notice that the phase $\Delta m \Delta t$ introduces automatically a time dependent relative
phase between the two amplitudes.  The marking and erasure operations can be performed on
entangled kaon pairs as in the optical case discussed above. The object kaon flying to the left
hand side is measured always \textit{actively} in the strangeness basis, see Fig.~\ref{QEkaon}(a).
As in the optical version the kaon flying to the right hand side, the meter kaon, is measured
\textit{actively} either in the strangeness basis by placing a piece of matter in the beam or in
the ``effective mass'' basis by removing the piece of matter. Both measurements are
\textit{actively} performed. In the latter case we obtain information about the lifetime, namely
\textit{which width} the object kaon has, and clearly no interference in the joint detections can
be observed.

\subsection*{(b) Partially passive quantum eraser with \textit{active} measurements}

\begin{figure}
\includegraphics[width=270pt,
keepaspectratio=true]{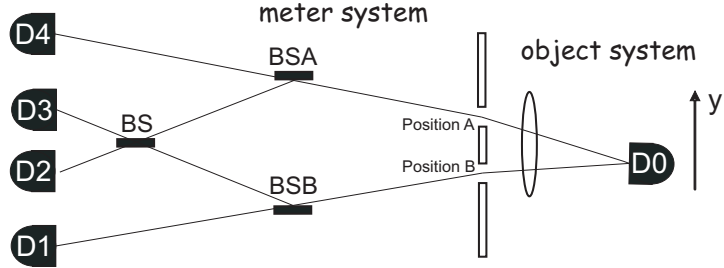} \caption{Here the setup of a partially active eraser is sketched.
An entangled photon pair can be produced either in region $A$ or in region $B$. If the detectors
$D1$ or $D2$ clicks, one knows the production region $A$ or $B$, i.e. one has full \textit{which
way} information. Clicks of the detectors $D2$ or $D3$ erase the information, interference is
observed. It is a partially active eraser, because the mark is erased by a probabilistic law,
however, the experimenter has still partially control over the erasure, she/he can choose the
ratio of transmittivity to reflectivity of the beam splitter $BSA$ and
$BSB$.}\label{photonPassive}
\end{figure}

In Fig.~\ref{photonPassive} a setup is sketched where either at position A or B an entangled
photon pair is produced, which was realized in Ref.~\cite{Kim}. ``Clicks'' on detector $D1$ or
$D4$ provide ``which way'' information. ``Clicks'' on detector $D2$ and $D3$ give no information
about the position A or B, interference is observed in the joint events of the two photons, see
Fig.~\ref{photonPassive}.

For kaons a piece of matter is permanently inserted into both beams where the one for the meter
system at the right hand side is fixed at time $t_r^0$, see Fig.~\ref{QEkaon}(b). The experiment
observes the region from the source to the piece of matter at the right hand side. In this way the
kaon moving to the right ---the meter system--- takes the choice to show ``which width''
information by its decay during its free propagation until $t_r^0$ or not by being absorbed in the
piece of matter. Again strangeness or lifetime is measured \textit{actively}. The choice whether
the ``wave--like'' property or the ``particle--like'' property is observed is naturally given by
the instability of the kaons. It is ``partially active'', because the experimenter can choose at
which fixed time $t_r^0$ the piece of matter is inserted. This is analogous to the optical case
where the experimenter can choose the transmittivity of the two beam--splitters $BSA$ and $BSB$ in
Fig.~\ref{photonPassive}.

Furthermore, note that it is not necessary to identify $K_S$ versus $K_L$ for demonstrating the
quantum marking and eraser principle.

\subsection*{(c) Passive eraser with ``\textit{passive}'' measurements on the meter}

Next we consider the setup in Fig.~\ref{QEkaon}(c). We take advantage ---and this is specific for
kaons--- of the \textit{passive} measurement. Again the strangeness content of the object system
---kaon moving to the left hand side--- is \textit{actively} measured by inserting a piece of matter into
the beam. In the beam of the meter no matter is placed in, the kaon moving to the right propagates
freely in space. This corresponds to a \textit{passive} measurement of either strangeness or
lifetime on the meter by recording the different decay modes of neutral kaons. If a semileptonic
decay mode is found, the strangeness content is measured. In the joint events interference is
observed. If a two or three $\pi$ decay is observed, the lifetime is observed and thus ``which
width'' information of the object system is obtained, no interference is seen in the joint events.
Clearly we have a completely passive erasing operation on the meter, the experimenter has no
control whether the lifetime mark is read out or not.

This experiment has no analog to any other considered two--level quantum system.

\subsection*{(d) Passive eraser with ``\textit{passive}'' measurements}

\begin{figure}
\begin{center}
\Large{{\textsf{\bf Kaonic quantum erasers}}}
\end{center}
\vspace{0.05cm}
\begin{flushleft}
(a) Active eraser with \textit{active} measurements (S: \textit{active}/\textit{active}; T: \textit{active})\\
\begin{center}\includegraphics[width=270pt, keepaspectratio=true]{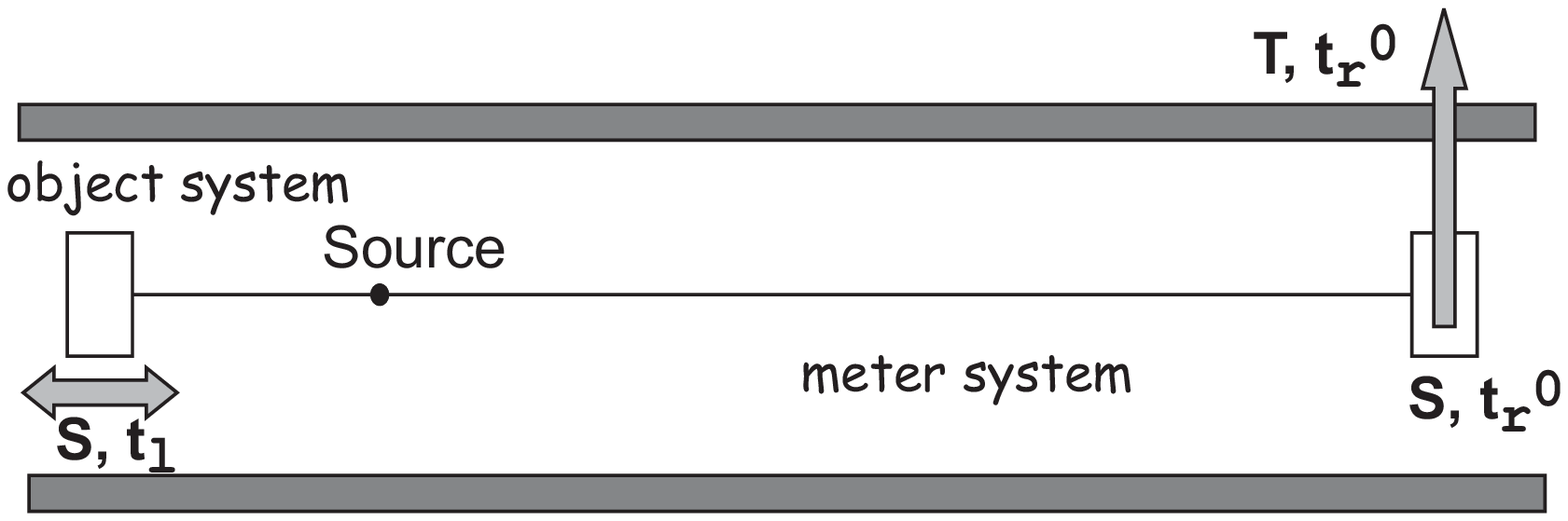}\end{center}
\vspace{0.4cm}
(b) Partially active eraser with \textit{active} measurements (S: \textit{active}/\textit{active}; T: \textit{active})\\
\vspace{0.3cm}
 \begin{center}\includegraphics[width=270pt, keepaspectratio=true]{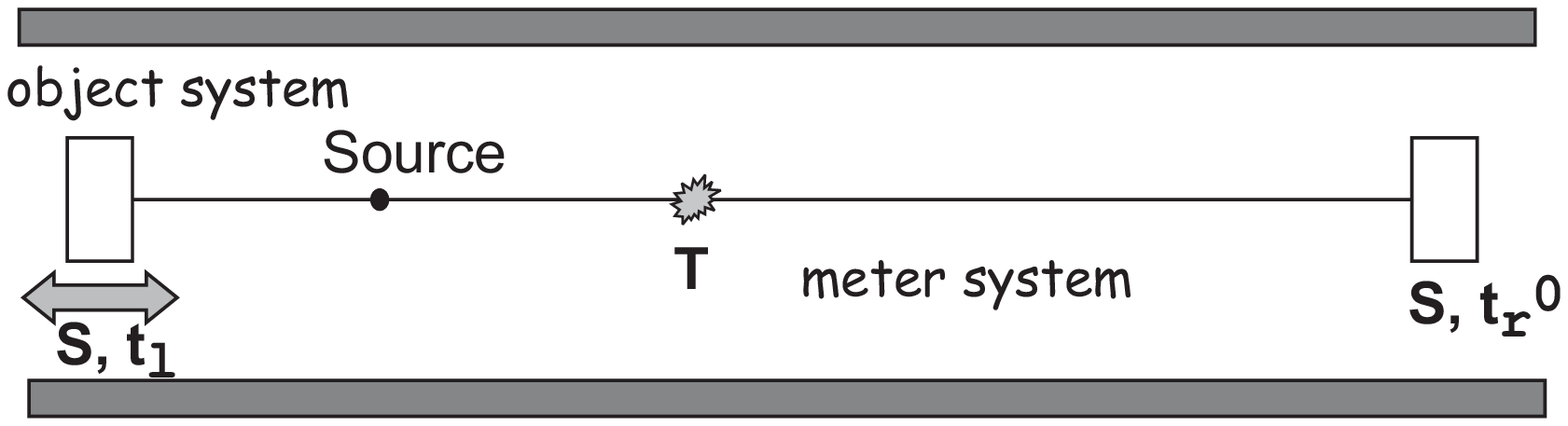}\end{center}
\vspace{0.4cm}
(c) Passive eraser with \textit{passive} measurements on the meter (S: \textit{active}/\textit{passive}; T:
\textit{passive})\\
\vspace{0.3cm}
 \begin{center}\includegraphics[width=270pt, keepaspectratio=true]{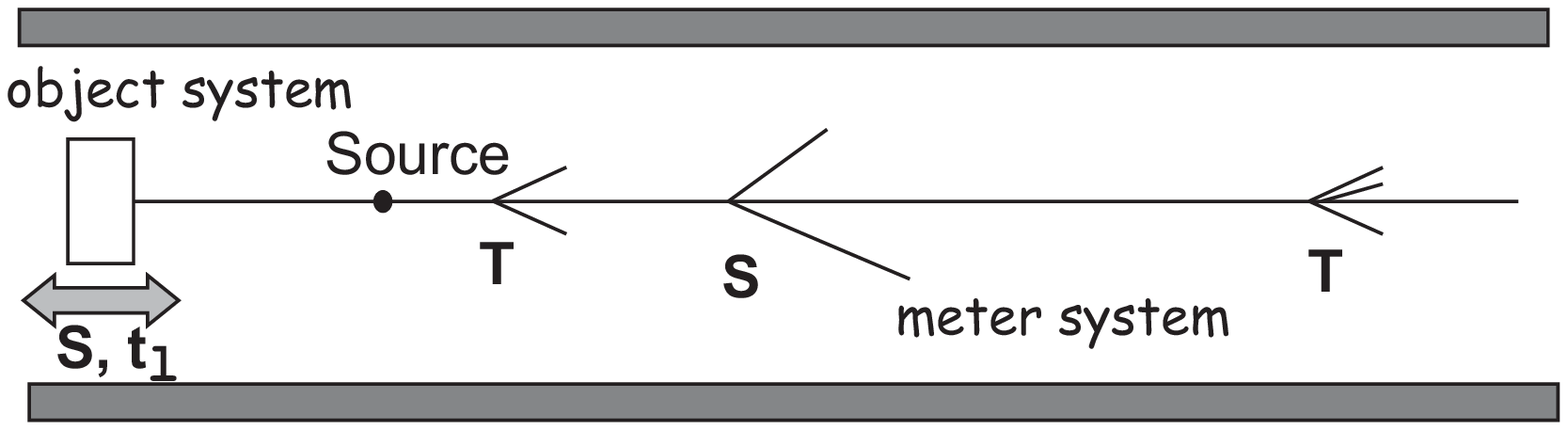}\end{center}
\vspace{0.4cm} (d) Passive eraser with \textit{passive} measurements (S:
\textit{passive}/\textit{passive}; T:
\textit{passive/passive})\\
\vspace{0.3cm}
\begin{center}\includegraphics[width=270pt, keepaspectratio=true]{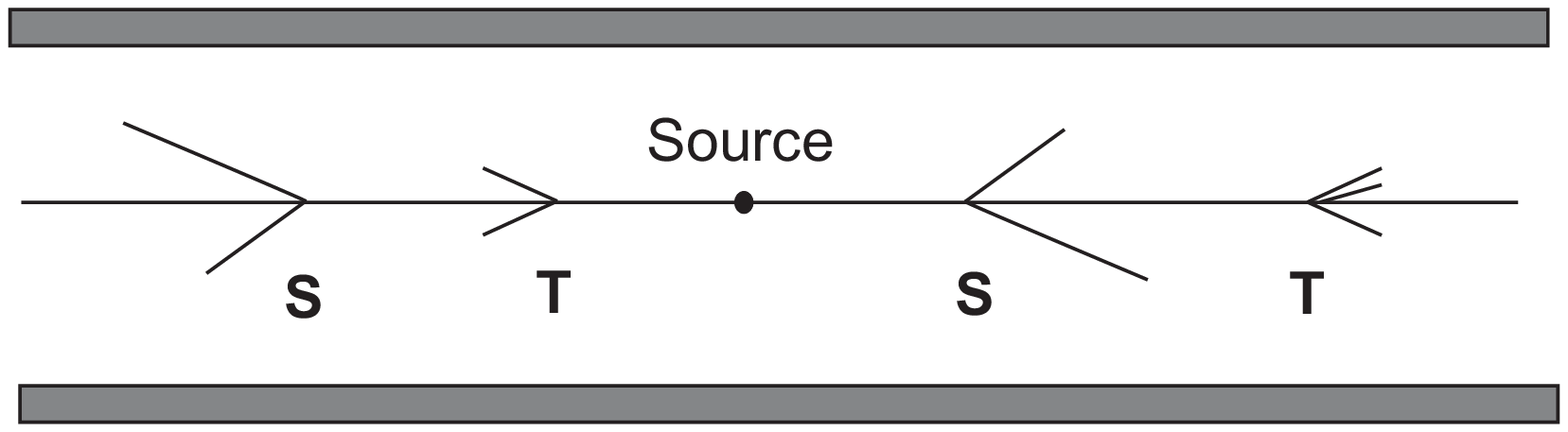}\end{center}
\caption{The figure shows four different setups for a quantum marking and quantum erasing
experiment. The first three, (a), (b) and (c), have the object system on the left hand side on
which the strangeness is always \textit{actively} measured at time $t_l$. The setups (a) and (b)
are analogous to existing quantum eraser experiments with entangled photons, see
Fig.~\ref{photonActive} and Fig.~\ref{photonPassive}. The setup (c) has no analog, because only
for kaons a \textit{passive} measurement is possible. For the last setup, (d), it is not so clear
which side plays the meter/object role as it is totally symmetric and it involves only
\emph{passive} measurements.}\label{QEkaon}
\end{flushleft}
\end{figure}

Finally we mention the setup in Fig.~\ref{QEkaon}(d), where both kaons evolve freely in space and
the experimenter observes \textit{passively} their decay modes and times. The experimenter has no
control over individual pairs neither on which of the two complementary observables at each kaon
is measured nor when it is measured. This setup is totally symmetric, thus it is not clear which
side plays the role of the meter and strictly speaking we cannot consider this experiment as a quantum eraser.\\

\vspace{0.3cm}

Summarizing, it is remarkable that for all four presented setups combining \textit{active} and
\textit{passive} measurement procedures lead to the same observable probabilities! This is even
true regardless of the temporal ordering of the measurements, thus kaonic erasers can also be
operated in the ``\textit{delayed choice}'' mode (for details see Ref.~\cite{SBGH6}).

\section{Conclusions}

In high energy physics neutral kaons ---concerning their properties as quasi--spin--- represent
qubits, which we call kaonic qubits. However, there are important differences in comparison to
photonic qubits. Kaons oscillate in strangeness, $K^0 \leftrightarrow \bar K^0$, they decay and
are characterized by $CP$ violation. Furthermore, kaon pairs also occur as entangled states in
analogy to entangled photon pairs.

We construct a Bell inequality for quasi--spins. To our surprise, the premises of local realistic
theories are \textit{only} compatible with strict $CP$ conservation in the $K^0 \bar K^0$ system.
In this sense $CP$ violation is a manifestation of the \textit{nonlocality} of the considered
state!

We also want to remark that the considered Bell inequality, since it is chosen at time $t=0$, it
is rather \textit{contextuality} than nonlocality which is tested. \textit{Noncontextuality}, the
independence of the value of an observable on the experimental context due to its predetermination
---a main hypothesis in hidden variable theories--- is definitely ruled out! So the contextual
quantum feature is demonstrated for entangled kaonic qubits.

However, kaons represent more than just qubit states. Due to the time evolution given by Nature
kaons can be considered as double slits corresponding to the two different decay states,
$|K_S\rangle , |K_L\rangle$. As time evolves both slits shrink, one faster than the other. The
contrast of the strangeness oscillation corresponds to the visibility of the interference pattern
and the ``which width'' information of the decay corresponds to the path predictability. With this
interpretation the validity of the quantitative complementarity relation is perfectly
demonstrated.

Finally, kaons are suitable systems to exhibit the amazing features of a quantum eraser. Compared
to photons we can study even more, namely ``active'' \textit{and} ``passive'' measurements, and
this offers the possibility to prove new eraser concepts. Four possible setups are constructed,
and remarkably all four ---the \textit{active} and \textit{passive} measurements--- lead to the
same observable probabilities. This illustrates nicely the very nature of a quantum eraser
experiment: it essentially sorts different events, namely, strangeness--strangeness or
strangeness--lifetime events representing the ``wave--like'' or the ``particle--like'' property.

Kaon experiments verifying the proposed quantum marking and eraser procedures have not been
performed till this day. Only the CPLEAR collaboration \cite{CPLEAR} did part of the job required
for the first setup of the active eraser.\\

{\bf Acknowledgement:} The authors acknowledge financial support from EURIDICE HPRN-CT-2002-00311.
We also would like to thank the referee for useful comments.



\begin{thebibliography}{100}


\bibitem{BertlmannSchladming}
R.A. Bertlmann, ``{\em Entanglement, {Bell} inequalities and decoherence in particle physics}'',
Lecture Notes in Physics (Springer-Verlag, Berlin, 2005), quant-ph/0410028.

\bibitem{BertlmannHiesmayr2001}
R. A. Bertlmann, B. C. Hiesmayr, Phys. Rev. A {\bf 63},  062112  (2001).

\bibitem{GisinGo}
N. Gisin and A. Go, Am. J. Phys. {\bf 69}, 264 (2001).

\bibitem{BBGH}
R.A. Bertlmann, A. Bramon, G. Garbarino and B.C. Hiesmayr, Phys. Lett. A \textbf{332}, 355 (2004).

\bibitem{Nagata}
K.~Nagata, W.~Laskowski, M.~Wie\' sniak and M.~\. Zukowski, Phys. Rev. Lett. \textbf{93}, 230403
(2004).

\bibitem{Unnikrishnan}
C.S.~Unnikrishnan, Europhys. Lett. \textbf{69}, 489 (2005).

\bibitem{Uchiyama}
F. Uchiyama, Phys. Lett. A \textbf{231}, 295 (1997).

\bibitem{BGH-CP}
R.A. Bertlmann, W. Grimus and B.C. Hiesmayr, Phys. Lett. A \textbf{289}, 21 (2001).

\bibitem{ParticleData}
D.E. Groom et al, \textit{Review of Particle Physics}, Eur. Phys. J. C \textbf{3}, 1 (1998).

\bibitem{Feynman}
R.P. Feynman, R.B. Leighton and M. Sands, \textit{The Feynman Lectures on Physics}, Vol. 3,
(Addison-Wesley, 1965), p. 1-1, p. 11-20.

\bibitem{GreenbergerYasin} D. Greenberger and A. Yasin, Phys. Lett. A {\bf 128}, 391 (1988).

\bibitem{Englert} B.-G. Englert, Phys. Rev. Lett. {\bf 77}, 2154 (1996).

\bibitem{SBGH3} A. Bramon, G. Garbarino and B. C. Hiesmayr,  Phys. Rev. A {\bf 69}, 022112 (2004).

\bibitem{HV}
B.C. Hiesmayr and V. Vedral, ``\textit{Interferometric wave-particle duality for thermodynamical
systems}'', quant-ph/0501015.

\bibitem{Arndt}
M. Arndt, O. Nairz, J. Vos-Andreae, C. Keller, G. Van der Zouw and A. Zeilinger, Nature {\bf 401},
680–682 (1999).

\bibitem{scully82}
M. O.~Scully and K.~Dr$\ddot{\rm u}$hl, Phys. Rev. A {\bf 25}, 2208 (1982).

\bibitem{Duerr}
S. D$\ddot{\rm u}$rr and G. Rempe, Opt. Commun. {\bf 179}, 323 (2000).

\bibitem{Herzog}
T.J. Herzog, P.G. Kwiat, H. Weinfurter and A. Zeilinger, Phys. Rev. Lett {\bf 75}, 3034 (1995).

\bibitem{Kim}
Y.-H. Kim, R. Yu, S.P. Kuklik, Y. Shih and M.O.~Scully, Phys. Rev. Lett. {\bf 84}, 1 (2000).

\bibitem{Tsegaye}
T. Tsegaye, G. Bj$\ddot{\rm o}$rk, M. Atat$\ddot{\rm u}$re, A.V. Sergienko, B.W.A. Saleh and M.C.
Teich, Phys. Rev. A {\bf 62}, 032106 (2000).

\bibitem{Walborn}
S.P. Walborn, M.O. Terra Cunha, S. Padua and C.H. Monken, Phys. Rev. A {\bf 65}, 033818 (2002).

\bibitem{Trifonov}
A. Trifonov, G. Bj$\ddot{\rm o}$rk, J. S$\ddot{\rm o}$derholm and T. Tsegaye, Eur. Phys. J. D {\bf
18}, 251 (2002).

\bibitem{KimKim}
H. Kim, J. Ko and T. Kim, Phys. Rev. A {\bf 67}, 054102 (2003).

\bibitem{AharonovZubaiy}
Y. Aharonov and M.S. Zubaiy, Science {\bf 307}, 875 (2005).

\bibitem{SBGH1} A. Bramon, G. Garbarino and B. C. Hiesmayr,
Phys. Rev. Lett. {\bf 92},  020405 (2004).

\bibitem{SBGH6} A. Bramon, G. Garbarino and B. C. Hiesmayr,
Phys. Rev. A {\bf 68},  062111 (2004).

\bibitem{CPLEAR}
A. Apstolakis et.al., Phys. Lett. B {\bf 422}, 339 (1998).


\end{thebibliography}
\end{document}